\documentclass[10pt, letterpaper]{article}
\usepackage{geometry}
\usepackage{cite}
\usepackage[utf8]{inputenc}
\usepackage{nameref,hyperref}
\usepackage[right]{lineno}
\usepackage{microtype}
\usepackage{amsmath}
\DisableLigatures[f]{encoding = *, family = * }
\raggedright
\makeatletter
\renewcommand{\@biblabel}[1]{\quad#1.}
\makeatother
\usepackage{graphicx}

\begin{document}

\begin{flushleft}
{\Large
\textbf\newline{Role of Poloidal $\mathbf{E}\times\mathbf{B}$ Drift in Divertor Heat Transport in DIII-D.}
}
\newline
\\
A.E. Jaervinen\textsuperscript{1,*},
S.L. Allen\textsuperscript{1},
A.W. Leonard\textsuperscript{2},
A.G. McLean\textsuperscript{1},
A.L. Moser\textsuperscript{2},
T.D. Rognlien\textsuperscript{1}, and
C.M. Samuell\textsuperscript{1}
\\
\bigskip
\bf{1} Lawrence Livermore National Laboratory, Livermore, CA 94550, USA
\\
\bf{2} General Atomics, San Diego, CA 92186, USA
\\
\bigskip
* jarvinena@fusion.gat.com

\end{flushleft}

\abstract{Simulations for DIII-D high confinement mode plasmas with the multifluid code UEDGE \cite{RognlienPoP1999} show a strong role of poloidal $\mathbf{E}\times\mathbf{B}$ drifts on divertor heat transport, challenging the paradigm of conduction limited scrape-off layer (SOL) transport. While simulations with reduced drift magnitude are well aligned with the assumption that electron heat conduction dominates the SOL heat transport, simulations with drifts predict that the poloidal convective $\mathbf{E}\times\mathbf{B}$ heat transport dominates over electron heat conduction in both attached and detached conditions. Since poloidal $\mathbf{E}\times\mathbf{B}$ flow propagates across magnetic field lines, poloidal transport with shallow magnetic pitch angles can reach values that are of the same order as would be provided by sonic flows parallel to the field lines. These flows can lead to strongly convection dominated divertor heat transport, increasing the poloidal volume of radiative power front, consistent with previous measurements at DIII-D \cite{LeonardPRL1997}. Due to these convective flows, the Lengyel integral approach \cite{LengyelIPP1981, GoldstonPPCF2017, PostJNM1995, KallenbachPPCF2013, ReinkeNF2017}, assuming zero convective fraction, is expected to provide a pessimistic estimate for radiative capability of impurities in the divertor. For the DIII-D simulations shown here, the Lengyel integral approach underestimates the radiated power by a factor of 6, indicating that for reliable DIII-D divertor power exhaust predictions, full 2D calculations, including drifts, would be necessary.}



\section{Introduction}

Heat transport along magnetic field lines in tokamak scrape-off layer (SOL) is commonly considered to be dominated by electron heat conduction. The reasons for this are clear. Due to the inverse of square root of mass dependence of the heat conductivity, ion heat conductivity is about 60 times lower than electron heat conductivity in deuterium plasmas. Neglecting heat carried by electrical currents, inertial heat flows, and ionization potential energy, the heat convection along the magnetic field lines is given by $5nTMc_s$, where for simplicity a pure deuterium plasmas is assumed with equal electron and ion densities, $n = n_e = n_i$, equal electron and ion temperatures, $T = T_e = T_i$, $M$ stands for the flow mach number, $M = v/c_s$, and $c_s$ is the plasma sound speed, $c_s \approx \sqrt{T/m_p}$, $m_p$ representing proton mass. Using this equation, we can calculate the equivalent Mach number that it takes to carry a given heat flux in the SOL, $M = q_\text{SOL}/5pc_{s}$, where $p$ stands for the static electron and ion pressure in the SOL. Assuming DIII-D scale parameters of $q_\text{SOL} \sim 0.5$ GW/m$^2$, $p_e \sim 450$ Pa \cite{LeonardNF2017}, we can calculate the $M\sqrt{T}$ that it takes to transport the heat convectively:
$M\sqrt{T} \approx 22.6 $ eV$^{0.5}$. Alternatively, assuming sound speed flow, $T = 510 $ eV, which is about a factor of 5 higher than the electron heat conduction limited maximum SOL temperature, $T \sim 120$ eV, at $q_\text{SOL} \sim 0.5$ GW/m$^2$. For ITER size plasmas, $q_\text{SOL} \sim $ a few GW/m$^2$, 3 GW/m$^2$ is assumed here, and $p_e \sim 1500 $ Pa, giving $M\sqrt{T} \approx 40.6 $  eV$^{0.5}$ or about $T = 1.6$ keV for sound speed flow along field lines, significantly higher than $T$ expected by electron heat conduction limited transport, $T \sim 200$ eV. Therefore, with assumptions made here, SOL heat transport along the field lines in these plasmas would be expected to be strongly dominated by electron heat conduction. 

However, cross-field drifts can convect heat flows across magnetic field lines. Since they transport particles and energy cross field lines, for shallow magnetic pitch angles, they can drive relatively strong poloidal transport even if their absolute magnitude is lower than the magnitude of parallel plasma flows. The equivalent parallel flow velocity for a poloidal cross-field flow can be calculated as $v_\parallel = v_\perp/\text{tan}(\alpha)$, where $\alpha = \text{arctan}(B_p/B_T)$, $B_p$ representing the poloidal field and $B_T$ the toroidal field. For magnetic pitch angles of 1 to 3 degrees, the corresponding multiplication factor is about 20 to 60. The magnetic drift flow velocities can be estimated as $v_{\nabla B} \approx 2T/(BR)$ \cite{RognlienPoP1999}. The equivalent parallel field line Mach number can be calculated as $M_{\nabla B} =  (2T/(BR))/(\text{tan}(\alpha) c_s)$. For a DIII-D size and field tokamak, $v_{\nabla B} \approx 10  - 30$ m/s, such that even for magnetic pitch angles of 1 degree, the effective $M_{\nabla B}$ is well below 0.05, which is significantly lower than sound speed flows expected at the target. For higher field and size tokamaks, such as ITER, this fraction would be even lower. 

On the other hand, electric drift velocities are calculated as $v_{\mathbf{E}\times\mathbf{B}} = E/B \propto 1/(\lambda_\Phi B)$, where $E$ stands for the electric field strength, $B$ for the total magnetic field strength, and $\lambda_\Phi$ for the electric potential scale length in the plasma. SOL electric fields of $\sim 5$ kV/m with $B_T \sim$ 2.1 T have been experimentally measured in DIII-D \cite{BoedoPoP2000}. These electric fields can drive cross-field flows of the order of 2.3 km/s, which, with magnetic pitch angles of 1 to 3 degrees, lead to equivalent parallel flow of 47 to 142 km/s. This is the equivalent of sound speed parallel flows at temperatures of 23 to 210 eV. Clearly electric drifts can drive SOL poloidal flows that correspond to the order of sonic flows parallel to field. The magnetic field in reactor scale tokamaks, such as ITER, is expected to be at least 2.5 to 3 times stronger than in DIII-D, such that the corresponding $v_{\mathbf{E}\times\mathbf{B}}$ may be expected to be lower as well. However, this is only the case if the $\lambda_\Phi$ does not reduce significantly with increasing magnetic field. The published scalings for heat flux width give approximately $\lambda_q \propto B_p^{-1}$ scaling \cite{EichNF2013}. Assuming fixed plasma shape, aspect ratio, and edge safety factor, $B_T \propto B_p$. If $\lambda_\Phi$ is assumed to be proportional to $\lambda_q$, then magnetic field dependence drops out from the scaling of $v_{\mathbf{E}\times\mathbf{B}}$. On the other hand, sound speeds in the divertor are expected to be the same in present day and next step tokamaks due to atomic physics constraining the operational divertor plasma temperature range. Therefore, even though scaling of the cross-field SOL scale lengths is beyond the focus of this paper and there may be important $\rho^*$ and SOL ballooning stability physics limiting the minimum scale lengths in the SOL \cite{ChangNF2017, EichNF2018}, it would be unjustified to declare cross-field drifts negligible for reactor scale tokamaks simply because their magnetic field is high.

In this paper, the role of $\mathbf{E}\times\mathbf{B}$ convection in divertor heat transport in DIII-D is analyzed and, the impact of this convection on the analytic radiated power estimated obtained with the Lengyel integral approach is discussed \cite{LengyelIPP1981}.The simulations investigated here were first introduced in a previous publication \cite{JaervinenNF2019}.

\section{UEDGE Simulations of Divertor Heat Transport in DIII-D}\label{DIIID}

\begin{figure}[!htb]
\centering
\includegraphics[width=0.8\textwidth]{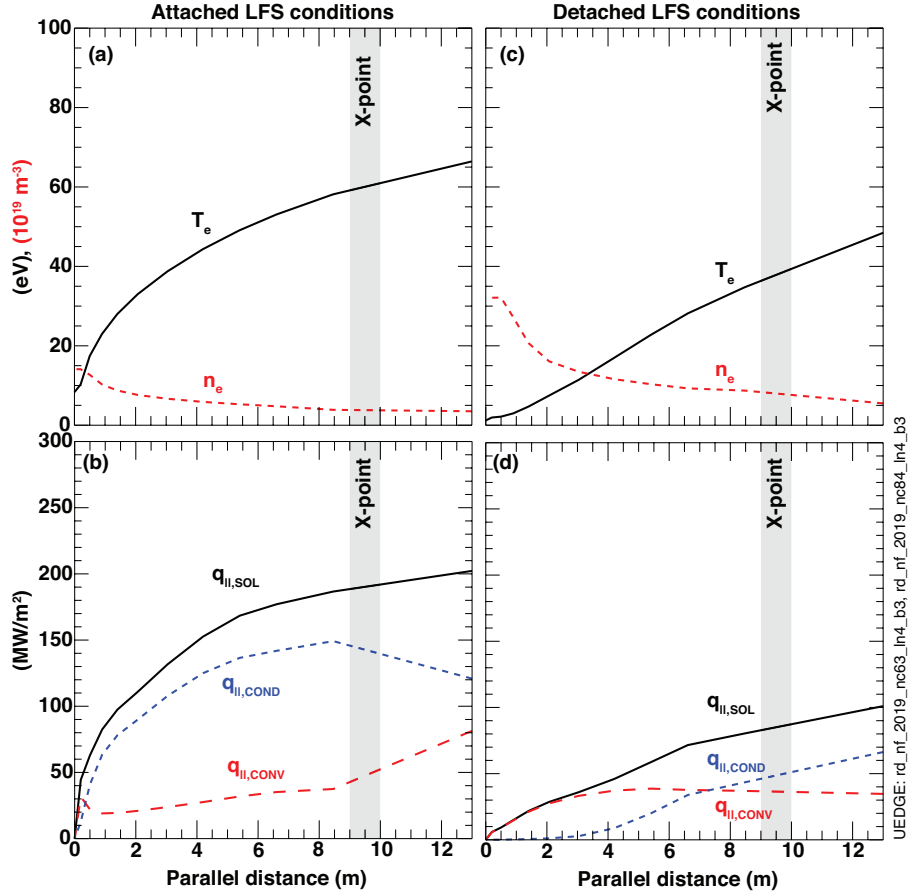}
\caption{Predicted low field side (LFS) divertor plasma electron temperature (black solid line) and density (red dashed line) in attached (a) and detached (c) LFS conditions with drifts multiplied by 1/3, as well as predicted total (black solid lide), conducted (blue dashed line), and convected (red dashed line) heat fluxes along the separatrix in LFS divertor in attached (b) and detached (d) LFS conditions. The x-axis represents parallel distance near the separatrix from the LFS target. The grey shaded area illustrates the poloidal location of the X-point.}\label{low_drift_figure}
\end{figure}

\begin{figure}[!htb]
\centering
\includegraphics[width=0.8\textwidth]{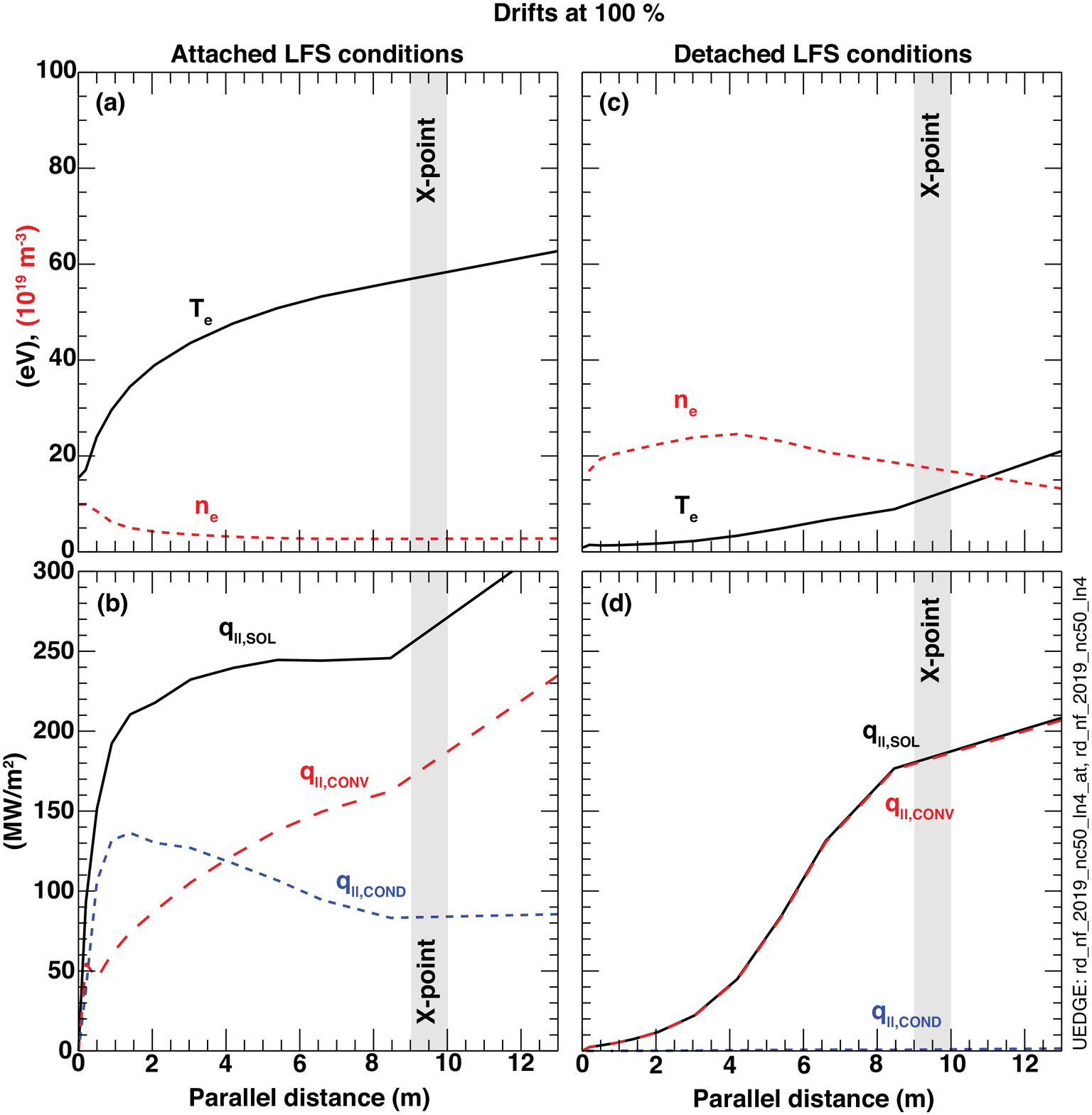}
\caption{Same profiles as in figure \ref{low_drift_figure} for UEDGE simulations with full drifts.}\label{normal_drift_figure}
\end{figure}

UEDGE simulations with cross-field drifts multiplied by a factor of 1/3 are well aligned with the conduction limited SOL heat transport paradigm (Fig. \ref{low_drift_figure}). In attached low field side (LFS) divertor conditions, with target $T_e \sim 10 $ eV, conducted heat flux carries about 80\% of the poloidal heat flux (Figs. \ref{low_drift_figure}a, b). In detached conditions, the fractional contribution of convected heat flux is predicted to increase as recycling driven flows take over at plasma temperatures below about 15 -- 20 eV (Figs. \ref{low_drift_figure}c, d). Overall, these simulations are well aligned with the paradigm that electron heat conduction is the dominant driver of heat transport in SOL with ion convection taking over near the divertor plate in strongly recycling or detached conditions.

Turning drifts fully on in these UEDGE simulations leads to a strong increase of the role of poloidal convection in the LFS divertor heat transport (Fig. \ref{normal_drift_figure}). In attached LFS conditions, convected heat flux carries about 50\% of the total SOL heat flux (Fig. \ref{normal_drift_figure}a, b), such that the dominance of electron heat conduction is already starting to seem questionable. In detached LFS conditions, the LFS divertor heat transport in the region where the dominant radiated power dissipation occurs, $T_e < 15$ eV, is predicted to be nearly 100\% dominated by convection, strongly in contrast with the conduction limited SOL heat transport paradigm (Fig. \ref{normal_drift_figure}c, d). Near the separatrix, the simulations are predicting radial electric fields of the order of about 5 to 7 kV/m, which is consistent with previous experimental measurements at DIII-D \cite{BoedoPoP2000}, although those experimental measurements were conducted in attached LFS divertor conditions. Conventionally, the radial electric potential gradients are thought to reduce in detached conditions as the radial sheath potential gradients are reduced with reducing plate temperatures. However, while this is the case near the target plate, the simulations show that upstream near the radiation front, strong radial potential gradients can be formed by other physics processes than the sheath potential drop, which is primarily driven by the radial $T_\text{e}$ gradient. Below the X-point, the residual radial component of the magnetic drift at the separatrix, which is of the order of 1 -- 2 m/s, is, with high plasma densities of 10$^{20}$ m$^{-3}$, sufficient to drive polarizing charge separation across the separatrix. With low temperatures and high resistivities in the divertor, this charge separation leads to potential hill formation in the PFR below the X-point, driving strong $\mathbf{E}\times\mathbf{B}$-flows below the X-point in detached conditions. These potential formation mechanisms are similar to those discussed by Rozhansky in \cite{RozhanskyCPP2018}. However, since the formed potential hill is not physically connected to wall structures, it leads to mostly self-closing $\mathbf{E}\times\mathbf{B}$ particle circulation around the potential hill. This private flux region (PFR) potential hill formation was previously discussed \cite{JaervinenPRL2018}. However, even though the particle flow loop is mostly self-closing, the heat carried with the particle flow is dissipated as the flow propagates through the divertor leg, such that for heat transport this circulation is not divergence free. 

Convective flow in detached conditions is strongly dominated by the $\mathbf{E}\times\mathbf{B}$ component in the simulations. The effective poloidal particle transport in the peak heat dissipation zone in the divertor leg equals to of the order of 4 -- 6 times sound speed flow along the field lines. The parallel flow Mach numbers are lower than 0.5 for most of the LFS divertor leg for all the simulated cases in this study. Clearly the $\mathbf{E}\times\mathbf{B}$-drift is transporting poloidal particle and heat flow that strongly exceeds the capability of the parallel fluid flow to carry convective transport. Since the poloidal $\mathbf{E}\times\mathbf{B}$-drift is carrying such a significant particle flux, the obvious follow up question is how does the particle balance work in the simulation. To satisfy particle continuity, flows are driven from source to sink and the flow loops must close such that the divergence of these flows is consistent with the source and sink profiles.  As was discussed previously, since the potential hill formation is not in contact with wall structures, the driven $\mathbf{E}\times\mathbf{B}$ circulation is mostly self-closing, such that the net particle transport is significantly lower than the gross particle transport and the divergence of the $\mathbf{E}\times\mathbf{B}$ flow is consistent with the particle source and sink profiles. Integrating over the the radial profile below the X-point from the PFR wall to about 2.5 mm into the common SOL, within which most of the $\mathbf{E}\times\mathbf{B}$ drift effects have a significant role, it is observed that about 74\% of the particle flux entering the divertor within the narrow $\mathbf{E}\times\mathbf{B}$-drift driven channel near the separatrix returns with the return flows. The integrated particle flux entering the LFS divertor within the area where drifts are significant is only about 26\% of the flux entering the divertor within the narrow $\mathbf{E}\times\mathbf{B}$-drift driven channel near the separatrix. However, for heat transport this is not the case as heat is dissipated from this particle flow pattern before it returns, such that the returning particles are cooler. As a result, for heat transport the return flow fraction is only about 23\% of the power entering the divertor within the peak heat flux area near the separatrix. 

\section{Impact of Drift Flows on Analytical Power Exhaust Calculations}\label{Lengyel}

The key impact of the strong convective fraction is that the parallel radiative volume in suitable electron temperatures for strong power exhaust is expanded. As a result, the Lengyel integral \cite{LengyelIPP1981, GoldstonPPCF2017, PostJNM1995, KallenbachPPCF2013, ReinkeNF2017} approach is expected to underestimate radiative capability of given impurity concentration in the divertor. In this approach, the upstream heat flux that can be dissipated by impurities in the SOL is calculated by integrating the radiated power density through the flux tube, while the heat transport in the flux tube is assumed to be carried by electron heat conduction. Furthermore, if plasma pressure is assumed to be constant in the flux tube the dissipated heat flux can be calculated as 
\begin{equation}
q_{\parallel, \text{det}} = nT \left(2 \int_{T_{\text{det}}}^{T_{\text{sep}}} f_z \kappa T^{0.5}L_z dT \right)^{0.5},
\end{equation}
where $f_z$ is the impurity concentration in the flux tube, $\kappa$ the electron heat conductivity divided by $T^{5/2}$, $L_z$ is the cooling rate coefficient for the given impurity, and $n=n_\text{e}$ and $T = T_\text{e}$ is used. In conduction limited divertor plasmas, this integral approach is expected to provide an appropriate approximation for the radiative capability of given impurity concentration. However, in conditions where significant convection occurs, the conduction limited assumption is expected to underestimate the electron temperature profile gradient scale lengths and, therefore, the parallel to field extent of the radiated power front. 

The difference between conduction limited and strongly convection dominated transport for radiated power can be as high as a factor of 6 in the examples shown here. To illustrate this further, a simple model is fitted to the radiation and plasma profiles in the detached case with full drifts. The model setup is as follows. $T \sim 1.0$ eV and $n \sim 2\times10^{20}$ m$^{-3}$ are given as boundary conditions at the target, as predicted by UEDGE. Target heat flux is obtained from these parameters, assuming sheath-heat transmission coefficient, $\gamma$ of 7: $q_{\parallel, \text{target}} = \gamma  n T c_s$. $\kappa \sim$ 1800 Wm$^{-1}$eV$^{-7/2}$ as approximately given by the UEDGE simulations here. Momentum losses in low $T$ are approximated by assuming that $n$ is constant below $T \sim 8$ eV. For $T > 8$ eV, usual static pressure balance is assumed: $nT = \text{constant}$.
Carbon concentration,$ f_z$, is set to 1.0\% close to values observed in the UEDGE simulations. Carbon cooling rates, $L(T)$, are calculated according to ADAS \cite{ADAS} data, assuming $n_e\tau = 10^{16}$ m$^{-3}$s, which gives cooling rates following the upper range of what is obtained with the predicted charge state distribution in the simulations. The calculated radiated power, $q_\text{rad} = n^2f_zL(T)$, is added to the parallel power flux when integrating the profiles upstream from the target. Finally a convective fraction of heat transport, $f_\text{conv}$, is given. The temperature gradient is calculated according to the electron heat conduction equation in all cases: $\nabla_\parallel T = (1 - f_\text{conv})q_\parallel/(-\kappa T^{5/2})$. This is justified as we constrain the convective power to be exactly a multiplier times the conductive power, which means that the effective Mach number is freely floating to ensure convective flow consistent with the specified $f_\text{conv}$. Due to this very simplifying assumption on the scaling of the convective flow, there is no predictive power in this simple model. It can only be used as a post processing tool to provide a simplified framework to illustrate some of the physics processes. Predicting the convective flow fraction requires a complicated, full 2D calculation, including drifts. 

\begin{figure}[!htb]
\centering
\includegraphics[width=0.95\textwidth]{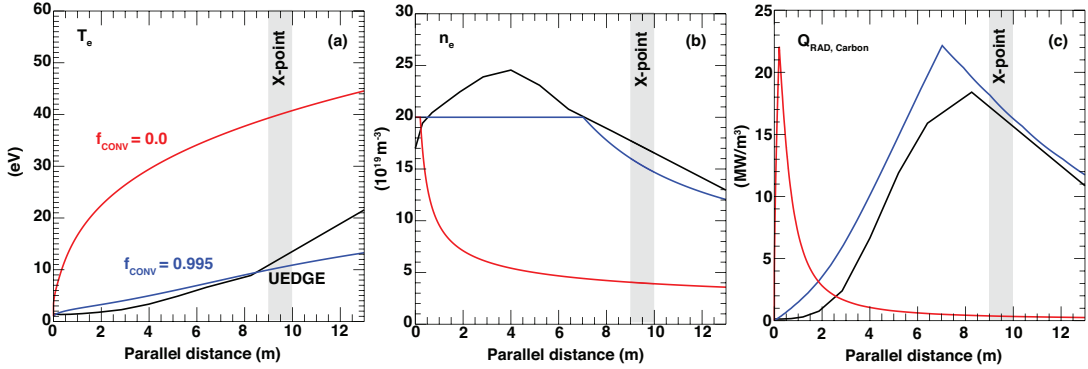}
\caption{Comparison UEDGE predictions (black) of LFS divertor $T_\text{e}$ (a), $n_\text{e}$ (b), and $Q_\text{RAD, Carbon}$ (c), as a function of parallel distance the LFS target near the separatrix to simple 1-D model with zero (red) and 99.5 \% (blue) convective fraction of heat transport.}\label{simple_model}
\end{figure}

Assuming a convective fraction of heat transport of 99.5\%, the simple model can be fitted to closely reproduce the UEDGE predictions for $T_\text{e}$ and carbon radiated power density, $Q_{\text{RAD, Carbon}}$ (Fig. \ref{simple_model}). Even though a good agreement is obtained between the simple model and the 2D simulations, it can only be obtained by incorporating physics information from the full 2D calculation, including drifts. There is very little predictive power in the simple model, but it can be used to illustrate how poorly the conduction dominated SOL assumption does in predicting the radiation profile. The overall dissipated heat by carbon radiation in the simple model is about 170 MW/m$^2$ close to values predicted by the UEDGE simulations. Assuming that the heat transport is 100\% conductive leads to steep electron temperature gradients and narrow radiation front, reducing the carbon radiation dissipation to about 28 MW/m$^2$, consistent with predictions that would be obtained with the Lengyel integral approach with these parameters (Fig. \ref{simple_model}). Therefore, the increased convective fraction of heat transport with the divertor drift flows can lead to a substantial, factor of about 6, increase of the radiative capability of low-Z impurities, such as carbon and nitrogen, that radiate in low temperatures of $T_\text{e} \sim 10$ eV, where heat conductivity is low. These predictions are consistent with previous experimental observations in the DIII-D tokamak \cite{LeonardPRL1997}, although the role of $\mathbf{E}\times\mathbf{B}$ drifts was not recognized in that previous study.

\section{Discussion}\label{discussion}

Analytic estimates for radiated power exhaust in the SOL with the Lengyel integral approach provide up to a factor of 6 less divertor radiation than calculated by UEDGE in highly convective conditions in detached DIII-D plasmas. These results clearly highlight that the simple analytic approach for predicting SOL conditions is prone to neglect physics processes that can lead to substantial changes in the predicted values. Simulations for DIII-D H-mode plasmas with the multifluid code UEDGE show a strong role of poloidal $\mathbf{E}\times\mathbf{B}$ drifts on divertor heat transport, challenging the paradigm of conduction limited SOL transport. Convective $\mathbf{E}\times\mathbf{B}$ flows are predicted to provide poloidal transport that is equivalent to poloidal transport provided by of the order of sonic particle flows parallel to the field. In detached conditions, radial potential gradients are formed in the divertor leg due to charge separation provided by radial magnetic drifts coupled with high electrical resistivity in low temperatures, similar to mechanisms discussed in \cite{RozhanskyCPP2018}. The resulting $\mathbf{E}\times\mathbf{B}$ flows can substantially increase the fraction of convective divertor heat transport, increasing the poloidal volume of radiative power front, consistent with previous measurements at DIII-D \cite{LeonardPRL1997}. Due to these convective flows, the Lengyel integral approach, assuming zero convective fraction, leads to a poor approximation of the poloidal extent of the radiation front. For the DIII-D simulations shown here, the Lengyel integral approach underestimates the radiated power density by a factor of 6. 

Even though the convective fraction of heat transport can be very high in DIII-D scale plasmas, the convective fraction is expected to reduce when scaling the physics to large size and higher $q_\text{SOL}$ facilities, such as ITER. The reason is that the heat convection for a given flow velocity is linearly proportional to the SOL pressure $\propto n_\text{e, SOL}T_\text{e, SOL}$. Sound speeds in the divertor are not expected to change, since the divertor temperatures are constrained by atomic physics requirements. Furthermore, upstream SOL temperature is not expected to increase strongly with $q_\text{SOL}$, $T_\text{e} \propto (q_\text{SOL})^{2/7}$. If the Greenwald limit scaling holds, SOL density is expected to be proportional to the Greenwald limit \cite{GreenwaldPPCF2002}, which is not expected to change significantly from DIII-D to ITER. Therefore, while the $q_\text{SOL}$ can be close to an order of magnitude higher in ITER than in the analyzed DIII-D plasmas here, the SOL pressure is expected to be only a factor of 3 -- 4 higher. Therefore, the heat convection fraction is expected to reduce. However, as the SOL electron temperature is reduced down to below 10 eV to approach detachment, it is expected that the heat flux in the SOL has been reduced also, such that the convective fraction can increase and potentially become significant near the detachment front formation in these next step facilities as well. Dedicated 2D simulations with drifts included would be required to address the role of these physics in next step devices. 

\section*{Acknowledgements}
This material is based upon work supported by the U.S. Department of Energy, Office of Science, Office of Fusion Energy Sciences, using the DIII-D National Fusion Facility, a DOE Office of Science user facility, under Awards DE-FC02-04ER54698, DE-AC52-07NA27344, and LLNL LDRD project 17-ERD-020.  
\textbf{Disclaimer:} This report was prepared as an account of work sponsored by an agency of the United States Government. Neither the United States Government nor any agency thereof, nor any of their employees, makes any warranty, express or implied, or assumes any legal liability or responsibility for the accuracy, completeness, or usefulness of any information, apparatus, product, or process disclosed, or represents that its use would not infringe privately owned rights. Reference herein to any specific commercial product, process, or service by trade name, trademark, manufacturer, or otherwise does not necessarily constitute or imply its endorsement, recommendation, or favoring by the United States Government or any agency thereof. The views and opinions of authors expressed herein do not necessarily state or reflect those of the United States Government or any agency thereof.

\providecommand{\url}[1]{\texttt{#1}}
\providecommand{\urlprefix}{}
\providecommand{\foreignlanguage}[2]{#2}
\providecommand{\Capitalize}[1]{\uppercase{#1}}
\providecommand{\capitalize}[1]{\expandafter\Capitalize#1}
\providecommand{\bibliographycite}[1]{\cite{#1}}
\providecommand{\bbland}{and}
\providecommand{\bblchap}{chap.}
\providecommand{\bblchapter}{chapter}
\providecommand{\bbletal}{et~al.}
\providecommand{\bbleditors}{editors}
\providecommand{\bbleds}{eds: }
\providecommand{\bbleditor}{editor}
\providecommand{\bbled}{ed.}
\providecommand{\bbledition}{edition}
\providecommand{\bbledn}{ed.}
\providecommand{\bbleidp}{page}
\providecommand{\bbleidpp}{pages}
\providecommand{\bblerratum}{erratum}
\providecommand{\bblin}{in}
\providecommand{\bblmthesis}{Master's thesis}
\providecommand{\bblno}{no.}
\providecommand{\bblnumber}{number}
\providecommand{\bblof}{of}
\providecommand{\bblpage}{page}
\providecommand{\bblpages}{pages}
\providecommand{\bblp}{p}
\providecommand{\bblphdthesis}{Ph.D. thesis}
\providecommand{\bblpp}{pp}
\providecommand{\bbltechrep}{}
\providecommand{\bbltechreport}{Technical Report}
\providecommand{\bblvolume}{volume}
\providecommand{\bblvol}{Vol.}
\providecommand{\bbljan}{January}
\providecommand{\bblfeb}{February}
\providecommand{\bblmar}{March}
\providecommand{\bblapr}{April}
\providecommand{\bblmay}{May}
\providecommand{\bbljun}{June}
\providecommand{\bbljul}{July}
\providecommand{\bblaug}{August}
\providecommand{\bblsep}{September}
\providecommand{\bbloct}{October}
\providecommand{\bblnov}{November}
\providecommand{\bbldec}{December}
\providecommand{\bblfirst}{First}
\providecommand{\bblfirsto}{1st}
\providecommand{\bblsecond}{Second}
\providecommand{\bblsecondo}{2nd}
\providecommand{\bblthird}{Third}
\providecommand{\bblthirdo}{3rd}
\providecommand{\bblfourth}{Fourth}
\providecommand{\bblfourtho}{4th}
\providecommand{\bblfifth}{Fifth}
\providecommand{\bblfiftho}{5th}
\providecommand{\bblst}{st}
\providecommand{\bblnd}{nd}
\providecommand{\bblrd}{rd}
\providecommand{\bblth}{th}

\end{document}